\documentclass{PoS}
\usepackage[numbers]{natbib}
\setcitestyle{number,open={[},close={]}}

\renewcommand{\deg}{$^{\circ}$}
\newcommand{\MV}{MultiView}
\newcommand{\uas}{$\mu$as}
\newcommand{\amin}{$^\prime$}
\def\mnras{MNRAS}

\def\aj{AJ}

\def\pasj{PASJ}

\title{Investigations on \MV\ VLBI for SKA}

\ShortTitle{\MV\ VLBI for SKA}

\author{\speaker{Richard Dodson}\\
        ICRAR/UWA, Australia\\
        E-mail: \email{richard.dodson@icrar.org}\\
        {Mar\'ia Rioja}\\
        ICRAR/UWA (Australia), CASS/CSIRO (Australia), OAN/IGN (Spain)\\
        E-mail: \email{maria.rioja@icrar.org}
        }

\abstract{The SKA will deliver orders of magnitude increases in sensitivity, but most astrometric VLBI observations are
limited by systematic errors. In these cases improved sensitivity offers no benefit. The best current solution for improving the accuracy of the VLBI calibration is \MV\ VLBI, where multiple simultaneous observations
around the target are used to deduce the corrections required for the line of sight to the target.
We have estimated and quantified the applicability of \MV\ from real-world ionospheric studies, making projections into achievable astrometric accuracies. These predict 
systematic measurement errors, with calibrators separated by several degrees, of $\sim$10\uas\ with current VLBI facilities. 
For closer calibrators, that are in-beam for single dish VLBI facilities, we predict systematic measurement errors of a few \uas. This is the ideal combination, where the sensitivity of the SKA will provide the precision and \MV\ will provide the accuracy.
Based on these results we suggest that the SKA design should increase the number of VLBI beams it can form from four to as many as ten. }

\FullConference{14th European VLBI Network Symposium \& Users Meeting (EVN 2018)\\
		8-11 October 2018\\
		Granada, Spain}

\begin{document}

\vspace*{-0.5cm}
\section{Introduction}
\vspace*{-0.3cm}

%#1
% SKA will have 100-times the collecting area of current telescopes.

% ⇒ Baselines to SKA will have 10-times the collecting area (√Aska Atel)

% SKA Mid and Low will be centred at frequencies around 1000 and 300 MHz, respectively
% ⇒ The new science will come at these frequencies

% Science targets will be newly discovered compact objects. 
% ⇒ VLBI will provide _dynamical_ information; the proper-motions, the relationship to other parts of the hosts, the distances 
% All astrometry — but astrometry at 1GHz and below is very hard

%#2
% Maria Rioja has covered the new methods for low frequency phase referencing. The errors arise from the static ionospheric component:
% Improvement come from:
% Reducing the ionosphere error,
% higher frequency, ensuring high Zenith angle,
% or reducing the source-calibrator separation.
% With dense GPS measurements we may be able to improve from a residual of 6TECU to 3TECU.
% But this is equivalent to a _metre_ of residual path length (20mm/o)
% c.f. 30mm of residual tropospheric path length (0.5mm/o)
% for 35μas we require ~1mm/o error on 6000km baseline
% ⇒ For significant improvements we need closer calibrators.

Very Long Baseline Interferometry (VLBI) can result in the highest angular resolutions achievable in astronomy and has a unique access to emission regions that are inaccessible with any other approach. 
%Therefore it holds the potential to increase our understanding of the physical processes in e.g. Active Galactic Nuclei (AGN), in the vicinity of super-massive Black Holes, and for studies of molecular transitions at high frequencies. 
One of the most powerful applications of VLBI is in the measurement of high precision relative astrometric angles between sources. Because  the reference sources are (usually) Quasars (QSOs) at high red shifts measurement of parallaxes with accuracies of $\sim$10\uas are possible, sufficient for measurement of distances across the span of our Galaxy. Phase referencing between sources allows for the precise correction of the atmospheric contributions, which are typically 3cm of wet delay from the troposphere and 6TECU of free-electron plasma from the ionosphere \citep[see, for example,][]{a07}.

Nevertheless the applications of astrometric-VLBI are not widespread outside of the centimeter wavelengths, say 5--22GHz. The observations become progressively more challenging away from these frequencies. 
%For the higher frequencies the system temperatures increase and the atmospheric coherence times fall and the sources are in general intrinsically weaker. 
For the lower frequencies the chief challenge is that the 
%recording bandwidth are smaller (as the central frequency approaches zero) and the 
ionospheric distortions are wildly different for different lines of sight. 
%We provide a summary of recent work that has been performed to extend the benefits achieved with conventional phase referencing, involving source switching, to hither-to unreachable regimes. These include the very high frequencies, i.e. mm and sub-mm VLBI where the short coherence times have prevented source switching, low frequencies, where direction dependent effects prevent source phase transfer, and decomposing the atmosphere into the Ionospheric and Tropospheric (in general, dispersive and non-dispersive) components and then correcting for these piece-wise. 

%We will discuss the implications for astrometry that arise from the parameters of the SKA: extremely high sensitivity at low frequencies. 
The SKA collecting area, and thus the sensitivity, will be two orders of magnitude larger than current facilities. 
Therefore when cross correlated with existing VLBI facilities there will be an order of magnitude increase in VLBI sensitivity. Furthermore when combined with the largest current VLBI facilities, such as 100m or above (e.g. the 500m FAST dish in China), it will be possible to achieve very close to the two orders of magnitude sensitivity gains.
The astrometry achievable with the thermal sensitivity limits from such an interferometer are of the order of 10 to 1\uas\ (i.e. a dynamic range of 1,000 to 10,000), depending on the size, number and configuration of the antennas. The challenge is to reduce the systematic errors to the same order, and this requires the development of innovative astrometric techniques that will address the underlying limitations for the SKA frequencies.
%Low frequency studies are largely driven by the efforts towards the building of the SKA array(s), which will demand new calibration solutions for the highly sensitive and low frequency baselines provided both by the array alone and in conjunction with other facilities. %The higher frequencies are driven by the revitalisation of mm-VLBI with the inclusion of ALMA and the excitement around the Event Horizon Telescope (EHT). 

\vspace*{-0.3cm}
\section{New Astrometric Techniques}
\vspace*{-0.3cm}

%Numerous indirect methods can be used to estimate the distance to young stars (e.g., de Grijs 2011), but they typically result in systematic uncertainties in excess of 20\%. Only trigonometric parallaxes can provide unbiased distance measurements, but they are notoriously challenging to obtain. For instance, the trigonometric parallax of a star at 200 pc is 5 milli-arcseconds (mas), so an astrometric accuracy of 50 micro-arcseconds (μas) on the parallax would be required to measure that distance to 1\% accuracy. This is more than one order of magnitude better than the astrometry delivered by the Hipparcos satellite (Perryman et al. 1997). Indeed, Hipparcos did not significantly improve our knowledge of the distance to star-forming regions in the Gould's Belt (e.g., Bertout et al. 1999). Also, the Hipparcos result on the distance to the Pleiades cluster, which is commonly used for testing theoretical stellar models, disagrees with all distance determinations obtained through other methods (Melis et al. 2014; David et al. 2016). The upcoming Gaia astrometric mission (de Bruijne 2012) will likely reach an accuracy of a few tens of μas, sufficient for percent accuracy determinations of distances in the Gould's Belt. However, since it operates at optical wavelengths, Gaia will be limited to stars that have low extinction. This will be an issue in star-forming regions like Orion, Ophiuchus, or Serpens, where values of AV larger than 10 are common (Cambrésy 1999; Ridge et al. 2006).

Current best-practise astrometric performance is currently at 22GHz (e.g. the results achieved in the BeSSeL project \citep{bessel_11}), where the troposphere dominates the astrometric budget. BeSSeL developed and makes extensive use of geodetic blocks to reduce the astrometric error from the static troposphere by a factor of three. A 1cm error on a 6000km baseline provides relative astrometric accuracy of $\sim$10\uas, with the calibrator at 1\deg\ separation.

The best 1.6GHz results come from finding very nearby calibrators; ionospheric residuals are typically 6TECU, which the equivalent of a position error of $\sim$100cm, at these frequencies. Close calibrators dilute the errors linearly as a function of separation (if the results are not dominated by thermal errors). Therefore, for 10\uas\ accuracy in the presence of 100cm errors, one would require a separation of about 1\amin. The PSRPi project \citep{deller_18} has demonstrated great success by finding calibration sources extremely close to the targets. These tend to be without well-known VLBI positions and weak (requiring a longer calibration solution interval). The median separation was 14\amin, with one out of 70 at 1\amin\ . The median per-epoch accuracy of these observations was about 100\uas. Note that the median final parallax accuracy (based on about 9 observations) was 60\uas, even for the closest pairing. In this case the weak calibrator source meant that the increased thermal noise dominated the very small systematic error. The PSRPi analysis was consistent with a quadratic combination of thermal errors plus a linear dependence on the separation, as in Asaki etal. \citet{a07}.

The only SKA requirements for conventional in-beam calibration are i) great sensitivity to allow for more (and therefore closer) calibrators for any arbitrary line of sight and ii) the ability to form two tied-array beams if the SKA tied-array beam size is smaller than the source separation, as is highly likely. But PSRPi found only one calibrator as close as 1\amin\  from the 70 PSR targets. As a SKA-VLBI baseline is the order of 10 times more sensitive we could only expect a source as close as 1\amin\  in $\sim$10\% of cases. This is inline with the analyses in Godfrey and Imai etal. (\citet{ska_135}, \citet{imai_ska}).

%We have examined the astrometric results from \citet{deller_18} from the PSRPi project. 70 pulsars were observed with conventional in-beam astrometric VLBI, with 60 having detectable in-beam sources. Of these the mean pulsar to calibrator separation was {\bf XX 10}', and the miniumum was {\bf 1'}. However the astrometric accuracy was limited by a combination of the closeness of the calibrator and the thermal noise. Assuming SKA-VLBI sensitivity is improved by an order of magnitude the fraction of sources detectable within a certain angular distance would also increase by close to an order of magnitude. There we might expect $\sim$10\% of SKA-VLBI targets to have an in-beam as close as 1', which would allow $\sim$10\uas\ astrometric accuracy. The term ($C$) for systematic errors given in the error analysis \citep[Sec 4.6]{deller_18} is consistent with those expected from nominal TEC residuals. 

\MV\ takes a new approach, by measuring the 2D phase surface over the antenna site \citep{rioja_17}. One measures the calibration terms in the direction of a number of calibrators around the target and projects those values to the actual line of sight to the target. These sources can then be both stronger and further from the target. The former allows for shorter solution intervals and smaller thermal errors. For a planar surface with perfect calibrators one would require 3 measurements to solve for the target direction. This was the basis of our previous recommendation for a minimum of 4 VLBI beams for SKA. We can think of no example where there is a benefit from taking calibrators outside the primary beam of the SKA antennas ($\sim$1deg) for SKA-VLBI. 
%Therefore sub-arraying (for astrometry) would not be a common mode. 
We focus on the case were the conventional VLBI antennas have the sources in-beam (i.e. a fraction of a degree for most cases) and the SKA ties all its stations together with multiple tied-array beams. We label this mode as `in-beam \MV '  as opposed to \MV\ with conventional VLBI switching.

% #3
% In-beam Phase Referencing addresses this directly:
%   e.g. PSR-π, which has typical separations of 0.2 degrees,
%   Possible for L-band, as usually find sources with-in VLBA beam;
%   PSR-π, 60 out of 70 sources had in-beams — high success rate 
% Rare for other frequencies as primary beam are smaller …
% Nevertheless for significant improvements we need even closer calibrators 
% SKA-VLBI will be an order of magnitude more sensitive: 
% so we are looking for a calibrator order of magnitude closer, searching an area two orders of magnitude smaller: 

% #4
% Nevertheless for significant improvements we need even closer calibrators 
% SKA-VLBI will be an order of magnitude more sensitive: 
% so we are looking for a calibrator order of magnitude closer, searching an area two orders of magnitude smaller: 
% All error terms will be zero (static/dynamic, tropo-/ionosphere)
% Perfect phase-referencing

% Demonstrated in Rioja ' 16 (visibility-based) & Reid ' 17 (image-based)
%     Used in Immer etal. 2018, Sakai etal. in-prep (virtual q\uasar)
%          solves for Static Ionospheric Wedge over array
         
\vspace*{-0.3cm}
\section{Crucial questions for SKA}
\vspace*{-0.3cm}
The crucial design considerations for VLBI-SKA, with respect to \MV\ astrometry are:\\
%\begin{itemize}
 \hspace*{1cm}$\bullet$ How many beams are needed to reduce the systematic errors to that of the thermal errors?\\
% \hspace{1cm}$\bullet$ Is it a function of frequency?
% \hspace{1cm}$\bullet$ Can we assume that the phase surface is flat?
 \hspace*{1cm}$\bullet$ Would more beams allow fitting a curved surface?\\
 \hspace*{1cm}$\bullet$ Would more beams allow contemporaneous checks on calibrators?
 %\hspace{1cm}$\bullet$ Would more beams allow new science goals?
%\end{itemize}

%Because of the limited resources in the beam-forming it is not possible to form an unlimited number of tied array beams. 
Initially the requirements were to form four beams, which is the minimum to measure calibrators in three directions, and to form a linear combination to solve for the target direction. 
But how reliable is that assumption, and is it sufficient for achieving the astrometric goal of 10\uas\ in a single epoch? The latter comes from the desire to match the systematic errors to the thermal errors in an observation with a SNR ratio of 1000, which would easily be achievable with VLBI observations between an SKA Phase-1 core and current VLBI resources.
Increasing the number of beams would allow for solving for non-linear ionospheric surfaces and for more careful cross checking of calibrators. 
Monitoring the latter to uncover unexpected changes in the reference position, which would be revealed by the shift of one particular calibrator against the others.
%In addition could more beams allow for new types of observations, such as surveys?
% #5
% These are the crucial design question for SKA-VLBI:

% How many beams are needed?
%               Is it a function of frequency?
% Can we assume that the phase surface is flat?
% Would more beams allow fitting a curved surface?
% Would more beams allow contemporaneous checks on calibrators?
% Would more beams allow new science goals?

% #6
% Dan Mitchell (Mitch) designed the Real Time System for EOR studies with Murchison Widefield Array (MWA). Chris Jordan used this to characterise MWA Phase-1 (3km baselines) ionospheric behaviour:
% Image-shift measurement for all visible sources, every 8-sec 

% Has been used to classify types of weather:
%  weak (1), moderately correlated (2),
%  highly correlated but weak (3), highly correlated and strong (4)
%\section{Sources of information}
To answer these questions we turned to two datasets, both from the Murchison Widefield Array (MWA). The first is the output from the Real Time System (RTS), which can be used to provide calibration information on the array with solutions every 8 seconds \citep{mitchell_08}. The 
%relevant data product here is the 
shift in the positions of known sources provides us with a measure of the average TEC gradient over the array.
%, in all directions that the array is sensitive to. For the MWA this is the majority of the visible sky, as each element is only 4m across. At 80MHz the first null is therefore $\pm$60\deg\ from zenith. 
The downside of this is that the average gradient over the array does not provide the fine scale structure of the TEC surface. For a 3km array (as for these data) and an ionospheric screen at a height of 300km the minimum angular scale is $\sim$0.6\deg. The results from RTS are most relevant to phase referencing with current VLBI arrays. 
We have taken 58 examples of the RTS dataset representing the four atmospheric conditions identified by Jordan etal. \citet{jordan_17}. This dataset contain one day each of: `weak' ionospheric structure (type 1), moderately correlated (type 2), highly correlated but weak (type 3) and highly correlated and strong (type 4). Jordan etal. \citet{jordan_17} provided an absolute characterisation of the ionospheric conditions. Our analysis was more focused on the relative differences, which is that required for relative phase referencing. 

For finer scale information, we turn to the by-product of the calibration method LEAP \citep{rioja_18}. This method provides the station-based calibration all visible calibrators, in parallel. LEAP offers an extremely promising solution for direction dependent (DD) calibration at low frequencies. It does not require a sky model, just calibrator positions, and therefore the DD calibrations are independent, allowing real-time and parallel solutions. The station-based solutions can be used to characterise the smoothness of the TEC screen at the sub-degree scale. This is the regime most relevant to (phase referencing with) the SKA (to in-beam sources for the single dish antennas). 
We have taken the calibration solutions from 18 Dec 2017, which has been classified as moderate weather.
%, and another day which was classed as unmanageable. 
All of these datasets were from the MWA Phase-2 array, with a maximum baseline of about 6km.

\vspace*{-0.4cm}
\subsection{RTS analysis}
\vspace*{-0.1cm}

\MV\ phase referencing will work perfectly if the TEC gradients (no matter how large) are the same in the different calibrator directions, because the 
%\MV\ calibration will work as long as the 
correction is linear. Therefore we are interested in the variance in the apparent shift in source positions as a function of angular separation, for all separations. We convert the measured shifts into the TEC gradient, which we express in TECU per degree. The mean value of the differences will capture curvature (the differential of the TEC gradients, $d\Delta{\rm TEC}$, over the FoV); for the variance we used the Root Mean Squared (RMS) value.
%RMS, as that (unlike standard deviation) does not subtract out the mean. 
%That is we form the following products, but the last is all that is relevant to the question:
\vspace{-0.5cm}
\begin{equation}
      d\Delta{\rm TEC}_{ij} \propto {\rm P}_i - {\rm P}_j ; 
      %d\Delta{\rm TEC}(\Omega) \propto \sum_{i\neq j} \overline{d\Delta{\rm TEC}_{ij}} ;
      \sigma\Delta{\rm TEC}(\Omega)) =  \sqrt{\sum_{i-j=\Omega}\overline{d\Delta{\rm TEC}_{ij}^2}}
\end{equation}
\vspace{-0.5cm}

where P$_i$ is the position offset of the $i^{\rm th}$ source and $\Omega$ is the angular separation between the $i^{\rm th}$ and $j^{\rm th}$ source. The 1D behaviour of the variance provides an estimate as to how far we can typically trust a \MV\ calibrator strategy for angular separations of one to tens of degrees.  

\vspace*{-0.1cm}
\subsection{LEAP analysis}
\vspace*{-0.1cm}

LEAP analysis is fully described in Rioja etal.  \citet{rioja_18}; here we limit ourselves to the analysis of the measured fine-scale TEC surfaces from LEAP. The $\Delta$TEC is measured for all calibrator directions across the field of view, after direction-independent and bandpass calibration. These individual TEC surfaces are a projection of the TEC onto the array and represent angular scales between the minimum and maximum baseline length (10--6000m) over the height of the ionosphere. We can use concurrent measurements of the ionospheric screen height from, for example the space weather service ({www.sws.gov.au}), or - with little loss of applicability - one can assume a canonical 300km. This gives angular scales 0.1 to 60\amin.
We assume that the measured TEC surface can be approximated by a planar surface (or a low-order polynomial) and measure the residuals. Assuming that the phase structure is a mixture of a simple structure plus turbulent noise we can discover i) how accurate a linear fit  would be in predicting the residual TEC in an arbitrary target direction and ii) how often a higher order fit would be a significant improvement in that prediction. Because it is vital to obtain results with the smallest possible intrinsic noise we limited our selves to the strongest  sources in each field of view. 
%We confirmed that we were not thermal noise limited by comparing with the results between 120s integration and 30s integration. As the noise did not double between these two datasets we could be sure that the solutions were neither dominated by thermal noise nor by fast changing solutions.

\vspace*{-0.5cm}
\section{Results}
\begin{figure}[htb!]
    \centering
    \includegraphics[width=0.24\textwidth]{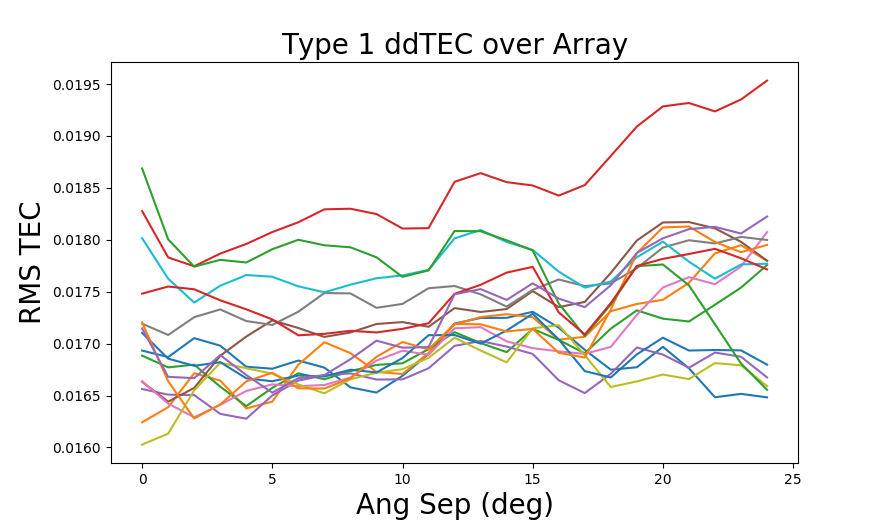}
    \includegraphics[width=0.24\textwidth]{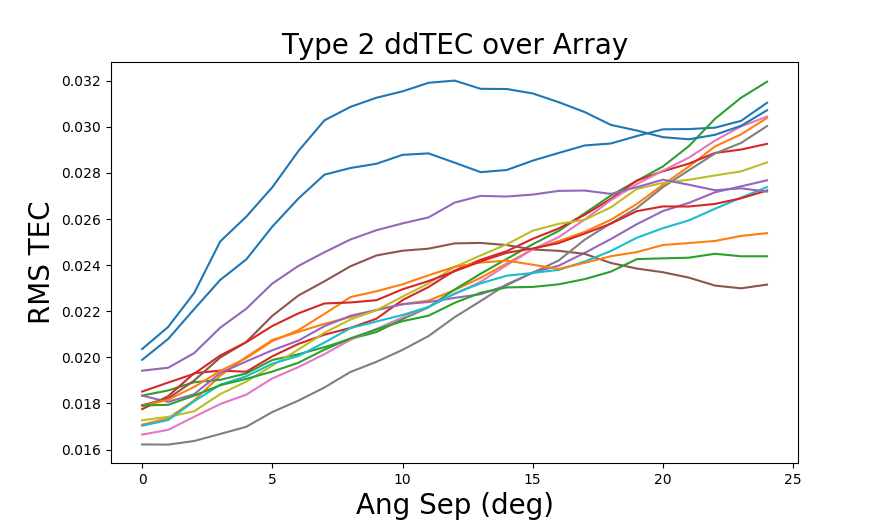}
    \includegraphics[width=0.24\textwidth]{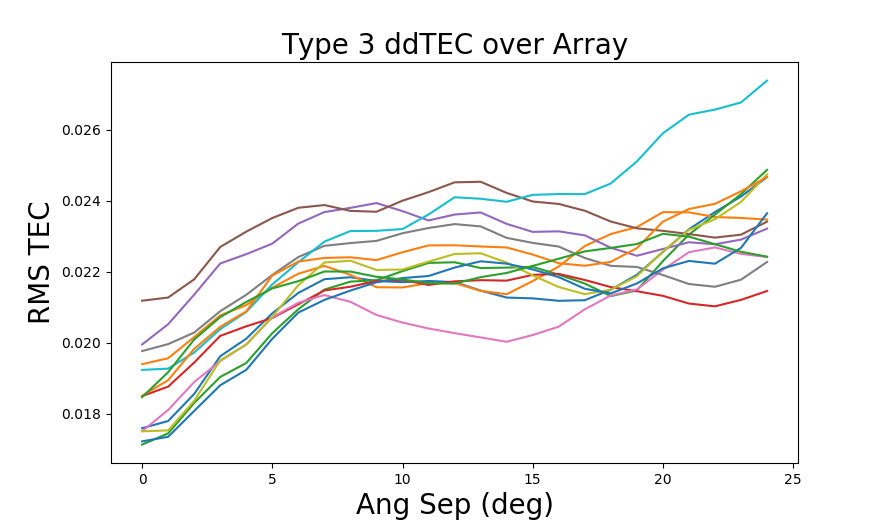}
    \includegraphics[width=0.24\textwidth]{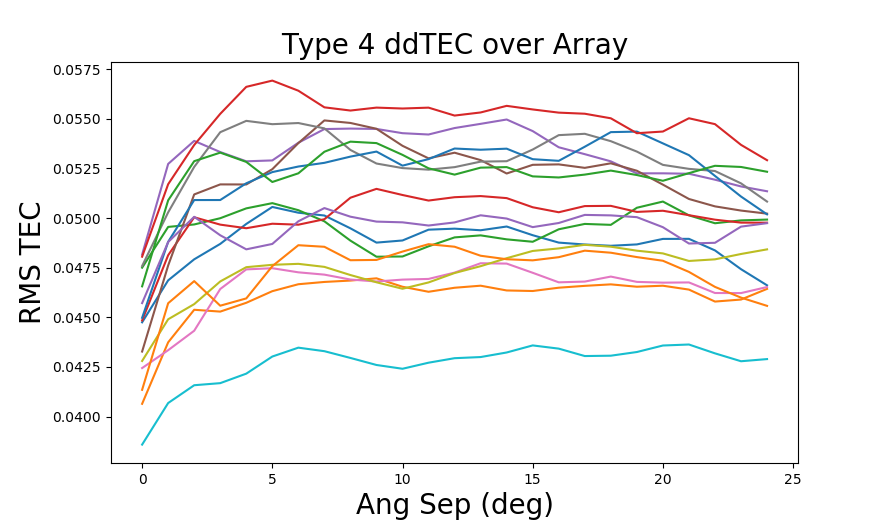}
    \caption{1-D projection of the variance (normalised by the separation) as a function of angular separation. The four classes identified in Jordan etal. \citet{jordan_17} are represented, running from type 1 to 4, left to right.}
    \label{fig:four_types}
\end{figure}

\vspace*{-0.3cm}
For the large angular scales we find that the four classes do have noticeably different behaviour, as shown in Fig. \ref{fig:four_types}. In type 1 the variance rose weakly with angular separation, with an average value of about 0.018TECU/\deg, in type 2 the variance rose significantly with angular separation, from $\sim$0.018 at small angular separations to nearly double that, at separations of 25\deg. Type 3 rose from from $\sim$0.018 at small angular separations and plateaued at about 0.022TECU/\deg\ between 5 and 20\deg. Type 4 rose rapidly between 1 and 2\deg\ to plateau at 0.05TECU/\deg. 

% #7
% In most cases (0.02dTEC/o) residual path at 1.5GHz is ~4mm for calibrators at 1o 
% ⇒ 100μas
% whereas, for BeSSeL-South (@6.7GHz)
% MV Calibrators with 3o  sep. would be acceptable in all weathers (0.05*3*400*6.7^-2)
% RTS Results are for LARGE SCALE >1o structure
% Applicable for switching or PAF VLBI
% \MV
% Will match in-beam at L-band with ~1o cals
% Will exceed in-beam above L-band

\begin{figure}
\begin{minipage}{0.44\textwidth}
\includegraphics[height=4cm,width=\textwidth]{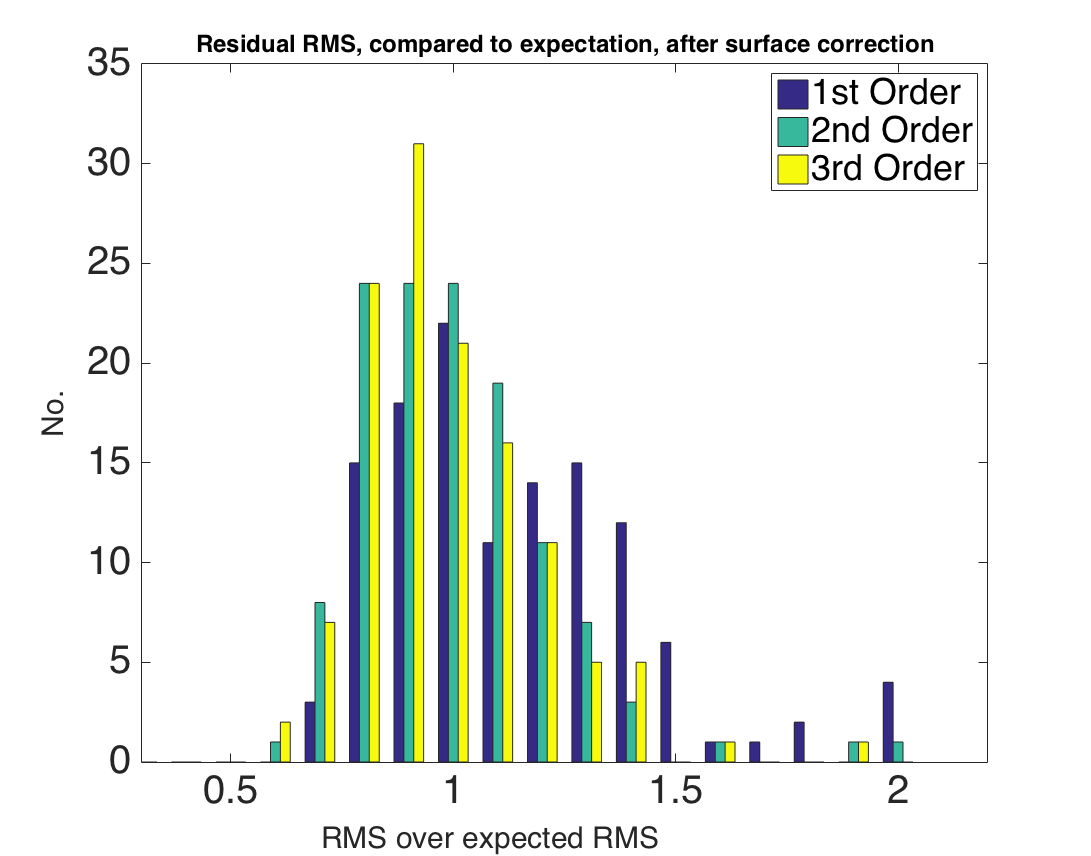}
\end{minipage}
\begin{minipage}{0.3\textwidth}
\caption{Deviation for the residual RMS from the expectation (based on the source flux), for 1st order, 2nd order and 3rd order fits. The 1st order fitting leaves 14\% of residuals greater than 3$\sigma$. \label{fig:hist}}
\end{minipage}
\begin{minipage}[c]{0.25\textwidth}
    \centering
        \begin{tabular}[c]{c||c|c|c}
        Order  & 1st & 2nd & 3rd \\
        \hline
        \% & 14 & 3 & 2\\
        \end{tabular}
        %\caption
        
        %\hspace{0.5cm}\vspace{0.3cm}
        {{\bf Table 1}: Fraction of RMS residuals greater than 3$\sigma$, as shown in Fig. 2. }
        %\label{tab:hist}
%\vspace{2.5cm}\\
\end{minipage}
\end{figure}

For the small angular scales we use the LEAP results to compare the residual RMS across the TEC (phase) surface against the expected value, given the source strength. Figure \ref{fig:hist} shows this distrubtion for 1st order (planar), 2nd order and 3rd order fitting. The distribution is clearly approximately Gaussian, with a mean close to one and $\sigma$ close to 0.14.
%0.2/$\sqrt{2}$. c**2=2sig**2
However 14\% of the planar fit residuals have a fractional RMS is higher than 3$\sigma$. See Table 1. These identify the TEC surfaces  with significant curvature.

%\vspace{-2cm}

\vspace*{-0.5cm}
\section{Conclusions}
\vspace*{-0.3cm}

At large angular scales ionospheric types 1, 2 and 3 would give acceptable astrometric results in \MV\ VLBI with current facilities, even at low frequencies. A TEC error of 0.02TECU is equivalent to 0.4cm at 1.4\,GHz. But note that these residuals are the imperfections from a planar surface, so the `dilution factor' of the angular separation do not apply. 
% \sigma_{res, cm} = 40*\Delta TEC \nu^{-2} = 0.4cm (per deg) at 1.4
% \Delta \theta_{us, MV} = \sigma_{res, cm}*1e-2 / 6000e3 *57.3 3600e6 = 140 
%        cf PR
% \sigma_{res, cm} = 40*\Delta TEC \nu^{-2} = 7cm w 6TECU at 6.7GHz
% \Delta \theta_{us, PR} = \sigma*1e-2 / 6000e3 * 3600e6 *57.3/57.3 = 40uas at 1 deg
Therefore even the type 3 behaviour (highly correlated but weak) would allow astrometry at the 140\uas\ level, for a set of calibrators with an average offset from the target of a degree. Obviously if the separations are double that, the errors are also doubled. 
This is compatible with the results from \citet{rioja_17}, where we found errors of $\sim$100\uas.
%In the worse case (type 4), the errors are approximately 2.5 times worse. But we note that even this would be sufficient to achieve the goal of 10\uas\ astrometric accuracy, per epoch. 
At higher frequencies, for example for 6.7GHz methanol masers (as being targeted by the BeSSeL project to extend the coverage of the galactic plane), we predict that even with angular separations of 5\deg\ in almost any conditions \MV\ accuracies would exceed those of observations with the conventional geodetic blocks.

Turning to the smaller angular scales, for the SKA in-beam \MV\ VLBI case, we find:
i) the source count estimates show the calibrator density will be insufficient to match the desired key-science goal of 10\uas, in 90\% of cases.
ii) That fitting planar surfaces give good solutions in 86\% of cases, with residuals as low as 2mTECU. % 5deg@150MHz
At this level one matches the potential thermal limits, which is equivalent to the ionospheric delay being $\sim$1mm (at 1.4GHz).
The remaining 14\% have significantly higher residuals, which can be reduced to the same level by fitting higher order polynomials. 
iii) At the level of 1\uas\ source stability will become a significant limitation, as this has only been demonstrated to $\sim$10\uas\ level \citep[e.g.][]{fomalont_11}. The only generally practical method to address this is to have more than the minimum number of calibrators.

% #12
% Astrometric requirements key driver for SKA-VLBI

% MWA measurements of SKA site phase screens
% show: range of ionospheric behaviours and classes
% suggest: acceptably linear over SKA-core 
% implies: excellent performance of in-beam \MV

% Suggested number of beams:
% 6 (minimum), 10 (lower goal) & 100 (maximum goal)

% Lower will lower systematic contributions to parallax to μas level
% Upper will allow deep phase referenced observations of every source in beam

We conclude that: 
In-beam astrometry will not achieve the target levels of 10\uas\ per epoch in  $\sim$90\% of cases;
Four beam \MV-VLBI will be able to reduce the errors from planar ionospheric contributions and allow us to achieve the target levels;
Four beam in-beam \MV-VLBI with the SKA-mid could achieve systematic errors of $\sim$1\uas, in $\sim$90\% of cases;
Calibrator instability will be a major source of error at this level of accuracy, requiring extra calibrators and beams;
Higher order fits to the TEC phase surface would allow systematic errors of $\sim$1\uas\ in nearly all cases.
%TEC resdiuals after \MV\ calibration would approximately follow $\Delta$TEC=2+20*$\Delta\theta$mTECU, where $\Delta\theta$ is the calibrator to target angular separation.

Therefore, to achieve the theoretical astrometric accuracy of the SKA-VLBI of $\sim$1\uas, the number of beams of SKA-VLBI should be increased from the current four to at least six, and potentially ten. This will provide robustness against the ultimate limit of non-linear ionospheric surfaces and/or calibrator instability.

{\bf Acknowledgements}: 
We thank C. Jordan for sharing the RTS datasets used in  the analysis.
%\end{acknowledgemnent}

%\begin{footnotesize}
% \bibliographystyle{plain}
% %\bibliographystyle{ksfh_nat} %EVN/JHEP}
% \bibliography{sfpr}  %referencias}

%\end{footnotesize}

\end{document}